\documentclass{optica-article}

\journal{opticajournal} 

\articletype{Research Article}

\usepackage{lineno}
\usepackage{subfig}
\usepackage[]{graphicx}
\usepackage{bm}

\usepackage{algorithm}      
\usepackage{algpseudocode}  

\begin{document}

\title{Modeling spectral filtering effects on color-matching functions: Implications for observer variability}

\author{Luvin Munish Ragoo\authormark{1,*}, Ivar Farup\authormark{1}, Casper F. Andersen\authormark{1} and Graham Finlayson\authormark{1,2}} 

\address{
\authormark{1}NTNU, Norway\\
\authormark{2}UEA, UK\\
}

\email{\authormark{*}luvin.m.ragoo@ntnu.no} 


\begin{abstract*} 
This study investigates the impact of spectral filtering on color-matching functions (CMFs) and its implications for observer variability modeling. We conducted color matching experiments with two  observers, both with and without a spectral filter in front of a bipartite field. Using a novel computational approach, we estimated the filter transmittance and transformation matrix necessary to convert unfiltered CMFs to filtered CMFs. Statistical analysis revealed good agreement between estimated and measured filter characteristics, particularly in central wavelength regions. Applying this methodology to compare between Stiles and Burch 1955 (SB1955)  mean observer CMFs and our previously published "ICVIO" mean observer CMFs, we identified a "\textit{yellow}" (short-wavelength suppressing) filter that effectively transforms between these datasets. This finding aligns with our hypothesis that observed differences between the CMF sets are attributable to age-related lens yellowing (average observer age: 49 years in ICVIO versus 30 years in SB1955). Our approach enables efficient representation of observer variability through a single filter rather than three separate functions, offering potentially reduced experimental overhead while maintaining accuracy in characterizing individual color vision differences.

\end{abstract*}


\section{Introduction}

Accurate color quantification is a cornerstone of color science, underpinning diverse applications such as colorimetry, color reproduction, material appearance perception, and color management. Central to this quantification are color-matching functions (CMFs), which describe the amounts of three reference stimuli required to match monochromatic test stimuli across the visible spectrum~\cite{Ohta_Robertson2005colorimetry}.

The Commission Internationale de l'Éclairage (CIE) has standardized CMFs by aggregating experimental data to represent the average color-matching responses of individuals with normal color vision~\cite{Schanda2007CIEColorimetry}. However, discrepancies between the predictions of standard CMFs and actual experimental matches made by color-normal observers have been widely documented~\cite{csuti2008colour,sarkar2011identification,asano2015individual,asano2016color}. These deviations highlight an essential feature of human color vision: individual variability.

Such variability arises from physiological differences in the human visual system, including variations in lens optical density~\cite{pokorny1987aging}, macular pigment density~\cite{smith1976variability}, genotypic differences in cone photopigments~\cite{thomas2011effect, stockman200042}, and age-related changes in ocular media~\cite{artigas2012spectral}. Understanding these individual differences has significant implications for industries reliant on precise color reproduction.

Efforts to characterize inter-observer variability in color-matching functions (CMFs) have taken different approaches. Asano et al.~\cite{asano2015individual, asano2016color} developed categorical models to classify observers based on their color-matching behavior, providing a framework for quantifying variability across individuals. Other studies, such as those by Shi et al.~\cite{shi2024multi}, have aimed to estimate physiological parameters—such as cone spectral sensitivities—that best fit an observer's color matches, revealing deviations from standardized models like the CIE 2006 observer~\cite{cie2006fundamental}. These approaches have advanced our understanding of inter-observer differences but primarily focus on modeling variability rather than exploring its underlying causes.

Andersen et al.~\cite{andersen2016estimating} tackled a different problem: estimating individual cone fundamentals from CMFs by combining lens and macular pigment effects into a single "pre-filter". This approach provided a way to relate individual CMFs to cone fundamentals through a linear optimization framework, offering insights into how optical filtering contributes to inter-observer variability. Inspired by this concept, our study aims to emulate the effect of Andersen's "pre-filter" experimentally by introducing a physical spectral filter into a color-matching setup.

Recent advances in computational filter design for imaging systems provide a framework for modeling transformations induced by spectral filters. Finlayson and Zhu~\cite{finlayson2019finding,FinlaysonZhuVora,FinlaysonZhu} developed methods to optimize spectral filters that transform camera spectral sensitivities to approximate the CIE XYZ color-matching functions. These methods employ alternating least-squares (ALS) optimization to minimize colorimetric errors while meeting physical constraints such as non-negative transmittance and smoothness. Integrating these filters with linear transformations has demonstrated significant improvements in imaging system accuracy~\cite{andersen2025}.

A key concept from these works is using pre-filters to modify a device's spectral characteristics, aligning them with standardized color representations. For instance, Finlayson and Zhu~\cite{finlayson2019finding} showed that an optimized filter placed in front of a camera can achieve near-perfect compliance with the Luther condition—a criterion ensuring that camera sensitivities are linear combinations of human CMFs. This principle illustrates how spectral filtering bridges device-specific responses and standardized models.

Building on these computational methods, our study extends their application to human vision by experimentally introducing a physical spectral filter into a color-matching setup. While Finlayson's work focuses on improving camera colorimetry, we investigate whether transformations derived from CMFs measured under filtered and unfiltered conditions can recover the physical properties of the filter itself. Aside from the recovery of the filter’s spectral characteristics from color-matching data, this approach also provides a framework to analyze how filter-induced transformations of the CMFs correspond to perceptual changes in human color vision resulting from spectral filtering.


\section{Current Study}

This study investigates the impact of an spectral filter on color matching behavior for two observers, Observer-I (male, age 52) and Observer-C (female, age 30). By maintaining consistent experimental conditions and varying only the filter's presence in front of the bipartite field, we aim to isolate its effect on color matching. This approach enables direct comparison of color-matching functions (CMFs) obtained with and without the filter, providing insights into the filter's influence on the observers' color matching process and potential reflections of inherent perceptual variations.

\subsection{Apparatus and Setup}
The experimental setup and method employed in this study are based on the apparatus and procedures detailed in our previous work~\cite{ragoo2024apparatus}, with a slight modification to accommodate the current experiment. The core component of the experimental apparatus is a custom-built colorimeter designed to measure individual color-matching functions (CMFs) through psychophysical experiments. This apparatus, constructed using 3D-printed integrating chambers, optical elements, and RGB LED light engines, creates a bipartite field allowing the observer to view two juxtaposed fields of light.

The test lights are near-spectral lights generated by a Bentham TMc300 monochromator, with peak wavelengths ranging from 400~nm to 720~nm at 10~nm intervals and a full width at half maximum (FWHM) of about 9~nm. A Bentham Il7 Xenon source serves as input to the monochromator. The bipartite field, with a diameter of 2~cm, is positioned approximately 80~cm from the observer, resulting in a visual angle of roughly 1.4$^\circ$.

The RGB light engine, which provides the reference lights, consists of a metal-core printed circuit board with four LEDs: two red (peak wavelength 634~nm), one green (534~nm), and one blue (452~nm). These LEDs yield the reference lights for the color-matching experiments. The RGB LED lights are controlled using PWM dimming and were set to run way below maximum intensity to minimize spectral shifts due to heating, ensuring precise and stable color reproduction throughout the experimental process.

For this specific experiment, the setup has been modified to include a "purple" filter that can be inserted in front of the bipartite field during the matching process. This filter is placed between the observer and the bipartite field for the color-matching procedure. Once a match is achieved, the filter is removed just before the spectral power distributions (SPDs) of the fields are measured. This modification allows for the comparison of color matches made with and without the filter, while ensuring that all SPD measurements are taken under consistent conditions.

\begin{figure}[ht]%
    \centering
    \subfloat[\centering Measured filter transmittance]{{\includegraphics[width=7cm]{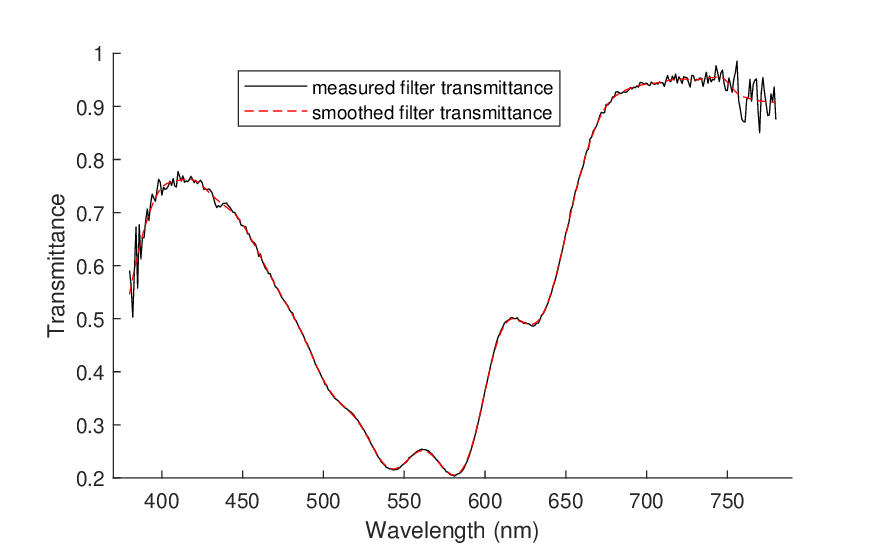} }}%
    \enspace
    \subfloat[\centering Filter transmittance resampled at every 10 nm]{{\includegraphics[width=6cm]{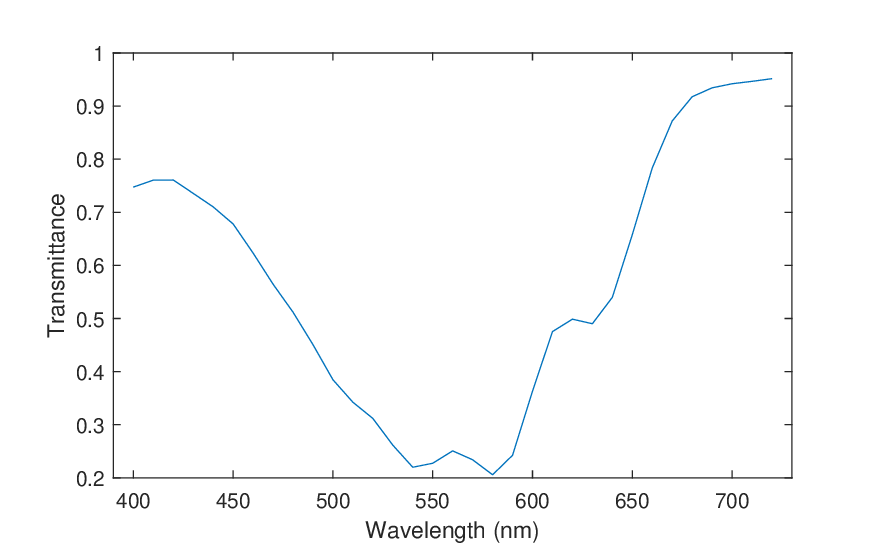} }}%
    \caption{Filter transmittance}%
    \label{Fig:FilterTransmittance}%
\end{figure}

The filter's transmittance was measured, smoothed and resampled at every 10~nm between 400~nm to 720~nm, as shown in Fig. \ref{Fig:FilterTransmittance}. 

The "purple" filter was specifically chosen because it attenuates the edges of the visible spectrum minimally, which is particularly advantageous for this experiment. At these edges, we already observe low luminance levels, as established in our previous work~\cite{ragoo2024apparatus}. By minimizing attenuation in these regions, the filter ensures that the inherent low luminance at spectral boundaries does not interfere with the color-matching process. This makes it highly compatible with the existing experimental setup, allowing for precise and reliable comparisons between matches made with and without the filter in place.

\subsection{Experimental Method}

The experiment was conducted in a dark room, with the bipartite field serving as the only light source. The observer's eyes were positioned approximately 80~cm from the bipartite field, corresponding to a visual angle of about 1.4°. This ensured that small head movements did not significantly impact color matching. Therefore, no head restraints were used, but the observer was instructed to loosely maintain the same viewing position.
The experiment began with a near-spectral test light of peak wavelength $\lambda_i$ set on the test side of the bipartite field, while the RGB LED lights, which in this experiment serve as reference lights, were initially off. The observer was instructed to freely scan the bipartite field with both eyes and adjust the intensities of the RGB reference lights to match the color of the reference field to that of the test field. In so doing, they were instructed to minimize differences in hue, brightness, and saturation between the two fields. Once a match was achieved, the SPDs of the test field, $\mathbf{T}_{\lambda_i}$ and the reference field, $\mathbf{M}_{\lambda_i}$ were measured using a spectroradiometer. Then, the reference lights were reset to zero, and a third measurement of the test light was taken by itself, $\mathbf{S}_{\lambda_i}$. This procedure was repeated for test lights with peak wavelengths ranging from 400~nm to 720~nm at every 10~nm interval. Several measurements (at least three, c.f. \cite{ragoo2024apparatus} for details) were made for each test light, to account for the observer's matching uncertainty. 
This experiment was repeated with the filter placed between the observer's eyes and the bipartite field, effectively integrating it as a transient component of the visual system during color matching. However, it is critical to note that only the perceptual matching process utilized the filter. Upon achieving a match, the filter was removed prior to any SPD measurements to ensure all radiometric data reflected the physical stimuli without spectral filtering. This approach isolates the filter's perceptual influence while preserving consistency in radiometric measurements for direct comparison between conditions.  

\subsection{Data processing}

\subsubsection{Computation of color-matching coefficients}

For each observer, two sets of raw color-matching data were obtained. One set corresponds to the experiment conducted without any filter, while the other set represents the case where a filter was placed in front of the bipartite field. The computation of tristimulus values followed the methodology outlined in our previous work~\cite{ragoo2024apparatus}. The process involved:
\begin{itemize}
    \item Scaling and normalization of the raw data to account for intensity variations and ensure comparability.
    \item Computation of color-matching coefficients using least-squares regression.
    \item Normalization of the coefficients to ensure positive values for reference lights and negative values for desaturation components. This step yields the tristimulus values at each test wavelength. 
\end{itemize}

\begin{figure}[ht]%
    \centering
    \subfloat[\centering without filter]{{\includegraphics[width=6cm]{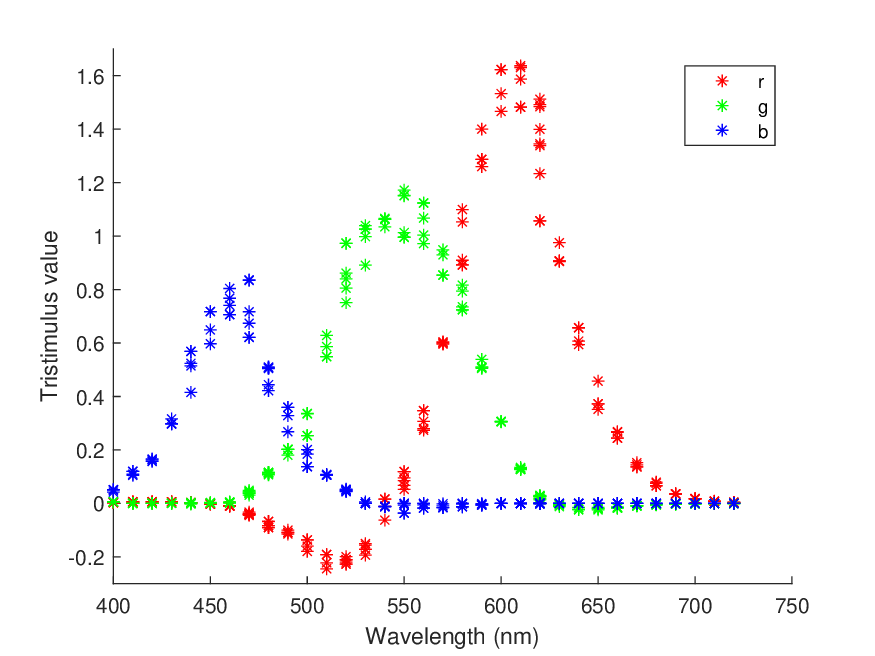} }}%
    \qquad
    \subfloat[\centering with filter]{{\includegraphics[width=6cm]{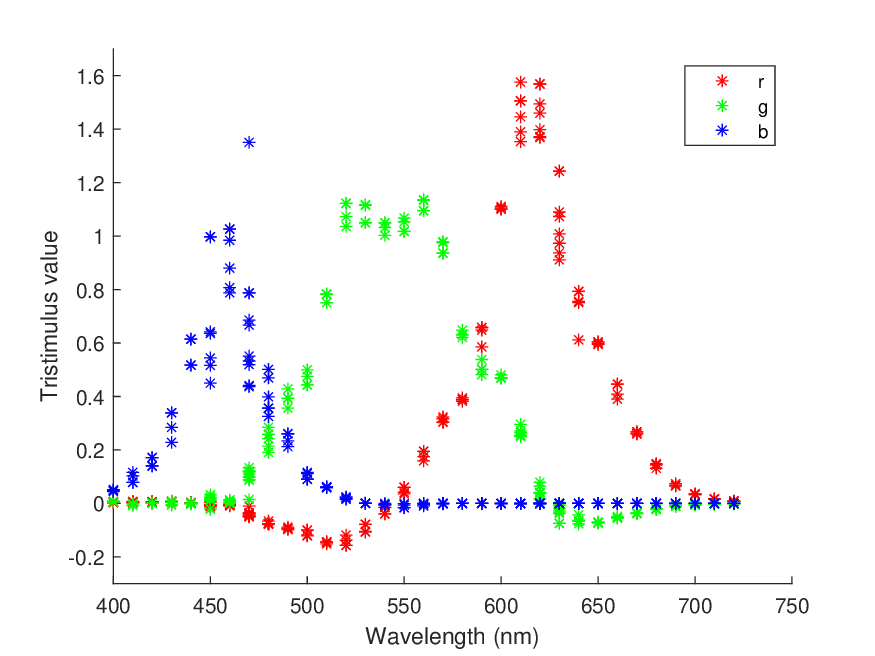} }}%
    \caption{Color-matching coefficients from color-matching experiment done (for Observer-I) : (a) without filter and (b) with filter}%
    \label{Fig:Color-matchingCoefficients}%
\end{figure}

The resulting tristimulus values (for Observer-I), expressed as functions of wavelength, are presented in Fig.~\ref{Fig:Color-matchingCoefficients} for both experimental conditions (with and without filter). This figure illustrates the tristimulus values relative to unit-energy normalized reference color stimuli \textbf{R}, \textbf{G}, and \textbf{B}, with peak wavelengths of 634~nm, 534~nm, and 452~nm, respectively. At each test wavelength, multiple data points are plotted, corresponding to the number of color matches made at that wavelength. This representation allows for a direct comparison of the filter's effect on the observer's color-matching behavior across the visible spectrum.

\subsubsection{Filter Estimation Algorithm}

The algorithm's goal is to estimate a matrix $\textbf{M}$ and a vector $\textbf{f}$ that would transform from the source CMFs (unfiltered condition), $\textbf{C}_{2}$, to the target CMFs (filtered condition), $\textbf{C}_{1}$ through: 

\begin{equation}
    \textbf{C}_1 = \mathrm{diag}(\textbf{f})  \textbf{C}_2 \textbf{M}^{-1}
\end{equation}

where:
\begin{itemize}
    \item $\textbf{C}_1, \textbf{C}_2 \in \mathbb{R}^{N \times 3}$: CMF matrices at 10~nm intervals (400-720nm) representing the target and source CMFs respectively. 
    \item $\textbf{M} \in \mathbb{R}^{3 \times 3}$: Transformation matrix
    \item $\textbf{f} \in \mathbb{R}^N$: Spectral transmittance vector (corresponding to "purple" filter's transmittance in experiment)
\end{itemize}

Essentially, the algorithm attempts to solve the optimization problem :

\begin{equation}
\label{Eq_optimisationProblem}
    \mathrm{argmin}_{\textbf{M},\textbf{f}} \left\|  \textbf{C}_1 \textbf{M} - \mathrm{diag}(\textbf{f})  \textbf{C}_2  \right\|^2
\end{equation}

subject to the constraint that the filter, 
 $\textbf{f}$, is non-negative and arbitrarily averages to unity in order to avoid ambiguity. 

 In order to stabilize the solution and avoid over fitting, a regularization term can be added to the optimization problem in Eq. \ref{Eq_optimisationProblem} as follows :

\begin{equation}
\label{Eq_optimProblem_regularised}
    \mathrm{argmin}_{\textbf{M},\textbf{f}} \left\|  \textbf{C}_1 \textbf{M} - \mathrm{diag}(\textbf{F})  \textbf{C}_2  \right\|^2 + \alpha \left| \textbf{T}  \textbf{f}\right|^2
\end{equation}

where :
\begin{itemize}
    \item $\textbf{T}$: First-derivative matrix enforcing spectral smoothness of the estimated filter:
    
    \begin{equation}
        \textbf{T} = \begin{bmatrix}
        1 & -1 & 0 & \cdots & 0 \\
        0 & 1 & -1 & \cdots & 0 \\
        \vdots & \ddots & \ddots & \ddots & \vdots
        \end{bmatrix} \in \mathbb{R}^{(N-1)\times N}
    \end{equation}
    
    \item $\alpha$: Regularization parameter (empirically determined)
\end{itemize}

This can be reformulated as a vector problem by introducing the matrices, $\textbf{A}$, $\Tilde{\textbf{T}}$  and vector $\mathbf{x}$ where,

\begin{align}
    \textbf{A} & = \underbrace{\begin{bmatrix}
        \textbf{C}_1 & \bm{0} & \bm{0} & -\operatorname{diag}(\textbf{C}_{2,1}) \\
        \bm{0} & \textbf{C}_1 & \bm{0} & -\operatorname{diag}(\textbf{C}_{2,2}) \\
        \bm{0} & \bm{0} & \textbf{C}_1 & -\operatorname{diag}(\textbf{C}_{2,3})
    \end{bmatrix}}_{ \mathbb{R}^{3N \times (9+N)}} \\
    \textbf{x} & = \underbrace{\begin{bmatrix}
        m_{11} & m_{21} & m_{31} & \dots & m_{33} & f_1 & f_2 & \dots  f_N
    \end{bmatrix}^T}_{\mathbb{R}^{9+N}}\\
    \Tilde{\mathbf{T}}  & = \underbrace{\begin{bmatrix}
        \bm{0} & \bm{0}  \\
        \bm{0} & \textbf{T}\\        
    \end{bmatrix}}_{\mathbb{R}^{(9+N) \times (9+N)}} \\
    \textbf{u} & = \underbrace{\begin{bmatrix}
        0 & \dots & 0 & 1 & \dots & 1   
    \end{bmatrix}}_{\mathbb{R}^{9+N}} \\
\end{align}

where $\textbf{C}_{2,i}$ refers to the $i^{th}$ column of $\textbf{C}_2$. \\

The regularised quadratic problem can be formulated as : 

\begin{equation}
\label{Eq_regQuadraticProblem}
    \begin{aligned}
     \textbf{x} & = \mathrm{argmin}_{\textbf{x}} \left(\left|  \textbf{A} \textbf{x} \right|^2 + \alpha \left| \Tilde{\textbf{T}}  \textbf{x}\right|^2\right)\\
     & = \mathrm{argmin}_{\textbf{x}}\textbf{x}^T\left(\textbf{A}^T\textbf{A} + \alpha\Tilde{\textbf{T}}^T\Tilde{\textbf{T}} 
     \right)\textbf{x}\\
     & = \mathrm{argmin}_{\textbf{x}}\textbf{x}^T \textbf{H}
     \textbf{x} \quad s.t. \quad \textbf{u}^T\textbf{x} = N
    \end{aligned}
\end{equation}

This optimisation problem can be solved using quadratic programming or Lagrangian multipliers~\cite{andersen2025}. The two terms in the optimization are both important. The $\textbf{x}$ minimizing 
$|\textbf{A} \textbf{x}|^2$ alone would be in the null-space of $A$, the dimension of which is $3N-9-N=2N-9$ (there would be many---in fact a whole space of---solutions). A {\it particular}  solution can be found thanks to the smoothness constraint $\alpha \left| \Tilde{\textbf{T}}  \textbf{x}\right|^2$.

\subsubsection{Estimating filter uncertainty}

The filter estimation algorithm aims to compute a spectral transmittance vector \textbf{f} (and a $3 \times 3$ transformation matrix), that transforms one set of CMFs (obtained without filter) into another set (obtained with a filter). However, a single estimate of \textbf{f} is insufficient for assessing the reliability of the algorithm, as it does not account for the inherent variability in the experimental measurements. This variability introduces uncertainty in the estimated filter, which reflects both experimental noise and potential inadequacies in the model itself. To address this, we leverage the multiple tristimulus measurements collected at each wavelength during the color-matching experiments to quantify uncertainty in the estimated filter. 

To estimate the uncertainty in \textbf{f}, we utilize a bootstrapping procedure that generates multiple realizations of CMFs based on the repeated measurements made during the experiments. This process allows us to compute an ensemble of filter estimates and subsequently perform statistical analysis. \\

\textbf{Bootstrapping procedure : }

\begin{enumerate}
    \item \textbf{Generating resampled CMFs : } 
    For each test wavelength, multiple tristimulus measurements were obtained during the experiments, both with and without the filter. By resampling these measurements with replacement, we generate "fictional" sets of CMFs for both conditions. Each resampled set represents a plausible variation of the observer's color-matching behavior under experimental noise. 
    \item \textbf{Filter estimation for each resampled Pair :}
    The filter estimation algorithm is applied to each pair of resampled CMFs (one with and one without the filter) to compute an independent estimate of the spectral transmittance vector. $\textbf{f}$. This results in an ensemble of filter estimates across all wavelengths. 
\end{enumerate}

The above steps are summarized in Algorithm \ref{Algorithm_FilterUncertaintyEstimation}.

\begin{algorithm}[H]
\caption{Bootstrapped Filter Estimation and Uncertainty Analysis}
\label{alg:bootstrapping}
\begin{algorithmic}[1]
\Require Repeated tristimulus measurements for CMFs without filter, $\textbf{C}_2$, and with filter, $\textbf{C}_1$, number of bootstrap iterations $N$, regularization parameter $\alpha$, smoothness matrix $\mathbf{T}$
\For{$i = 1$ \textbf{to} $N$}
    \State \textbf{Resample Data:} Randomly sample with replacement to obtain resampled CMFs:
    \[
    \mathbf{C}_1^{(i)} \text{ and } \mathbf{C}_2^{(i)}
    \]
    \State \textbf{Filter Estimation:} Compute filter estimate $\hat{\textbf{f}}^{(i)}$ by solving:
    \[
    \arg\min_{\hat{\textbf{f}}^{(i)}} \left\| \mathbf{C}_1^{(i)}\mathbf{M} - \operatorname{diag}(\hat{\textbf{f}}^{(i)}) \, \mathbf{C}_2^{(i)}  \right\|^2 
    + \alpha | \mathbf{T} \hat{\textbf{f}}^{(i)} |^2,
    \]

    \State \textbf{Storing Filter estimates:} Store $\hat{\mathbf{f}}^{(i)}$.
\EndFor

\end{algorithmic}
\label{Algorithm_FilterUncertaintyEstimation}
\end{algorithm}

For each wavelength $\lambda$, the mean filter estimate $\bar{\mathbf{f}}(\lambda)$ and its standard deviation $\sigma_{\mathbf{f}}(\lambda)$ are computed from the set of bootstrap filter estimates $\{\hat{\mathbf{f}}^{(i)}(\lambda)\}_{i=1}^N$ as follows:
\begin{equation}
    \bar{\mathbf{f}}(\lambda) = \frac{1}{N} \sum_{i=1}^{N} \hat{\mathbf{f}}^{(i)}(\lambda)
\end{equation}
\begin{equation}
        \sigma_\mathbf{f}(\lambda) = \sqrt{\frac{1}{N-1} \sum_{i=1}^{N} \left(\hat{\mathbf{f}}^{(i)}(\lambda) - \bar{\mathbf{f}}(\lambda)\right)^2}
\end{equation}



\section{Results}

\subsection{Mean color-matching functions (CMFs)}

\begin{figure}[h]%
    \centering
    \subfloat[\centering Experimental condition : no filter]{{\includegraphics[width=6.5cm]{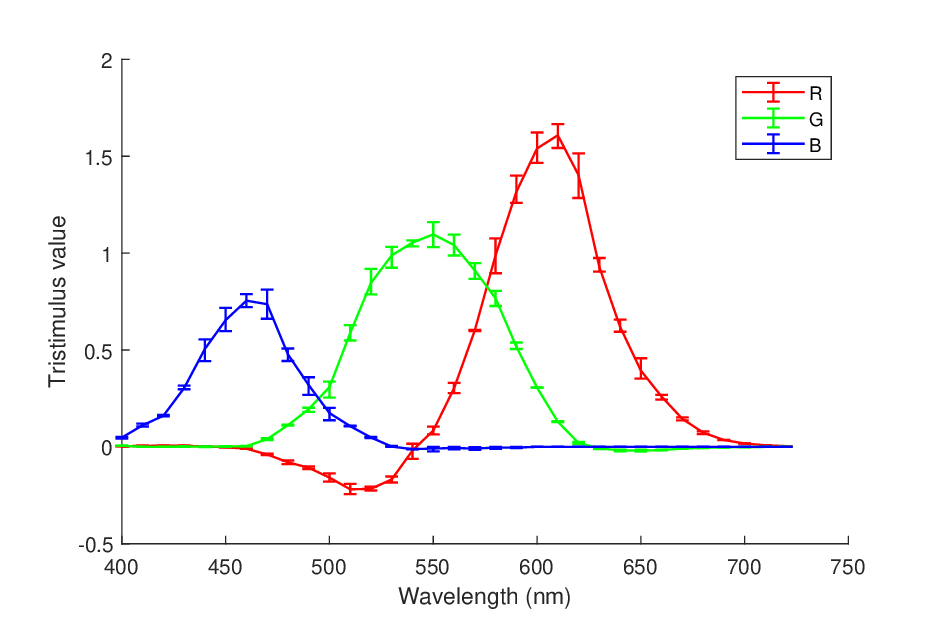} }}%
    \enspace
    \subfloat[\centering Experimental condition : with filter]{{\includegraphics[width=6.5cm]{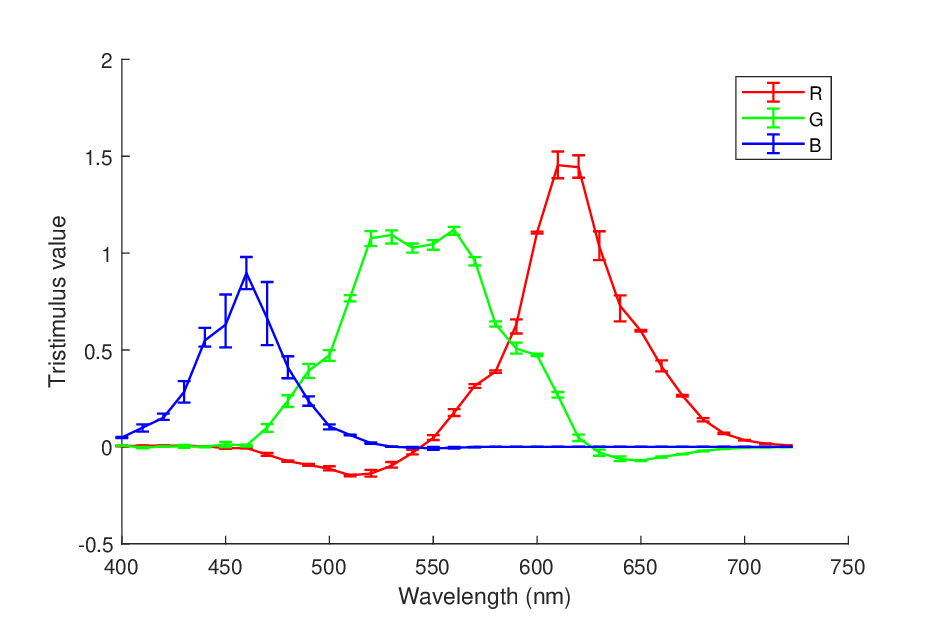} }}%
    \caption{Observer-I : Mean CMFs with error bars representing 95\% CI of the mean.}%
    \label{Fig:MeanCMFs_wErrorBars}%
\end{figure}

\begin{figure}[h]%
    \centering
    \subfloat[\centering Experimental condition : no filter]{{\includegraphics[width=6.5cm]{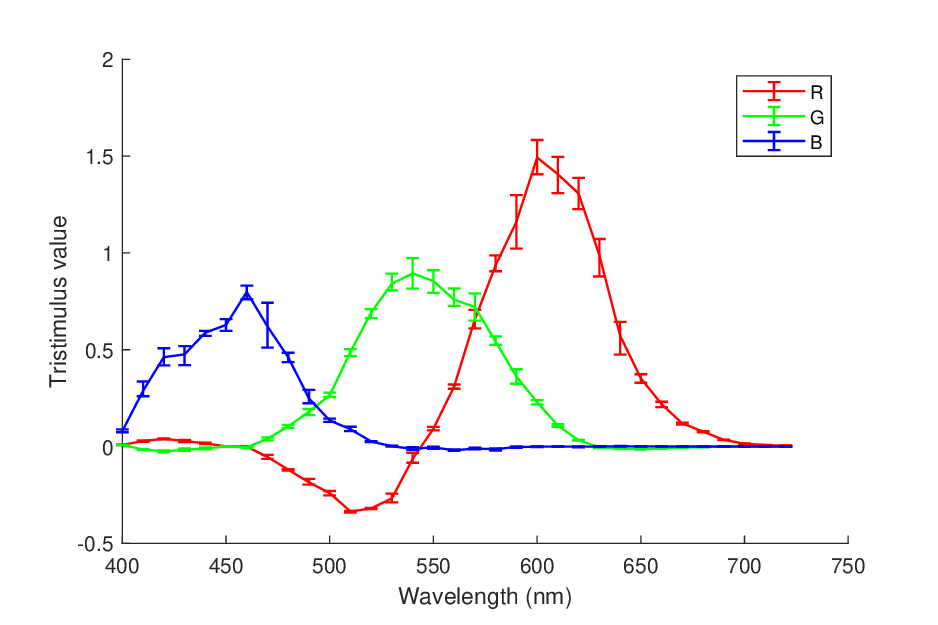} }}%
    \enspace
    \subfloat[\centering Experimental condition : with filter]{{\includegraphics[width=6.5cm]{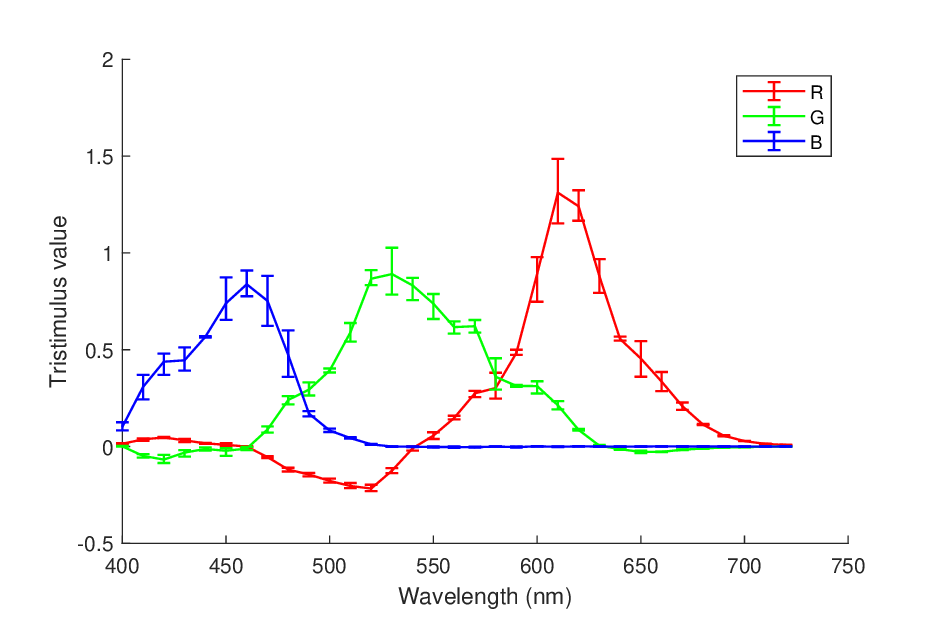} }}%
    \caption{Observer-C : Mean CMFs with error bars representing 95\% CI of the mean.}%
    \label{Fig:MeanCMFs_wErrorBars_Cristina}%
\end{figure}

The mean CMFs for the two observers under both experimental conditions, with and without the "purple" filter, are illustrated in Fig.~\ref{Fig:MeanCMFs_wErrorBars} and Fig.~\ref{Fig:MeanCMFs_wErrorBars_Cristina}. The CMFs are plotted as functions of wavelength, with error bars representing the 95~\% bootstrapped confidence intervals (CI) of the mean. This figure provides a visual comparison of how the presence of the filter affects each observer's color-matching behavior across the visible spectrum.

\subsection{Filter estimation and uncertainty analysis}

\begin{figure}[ht]
\centering\includegraphics[width=10cm]{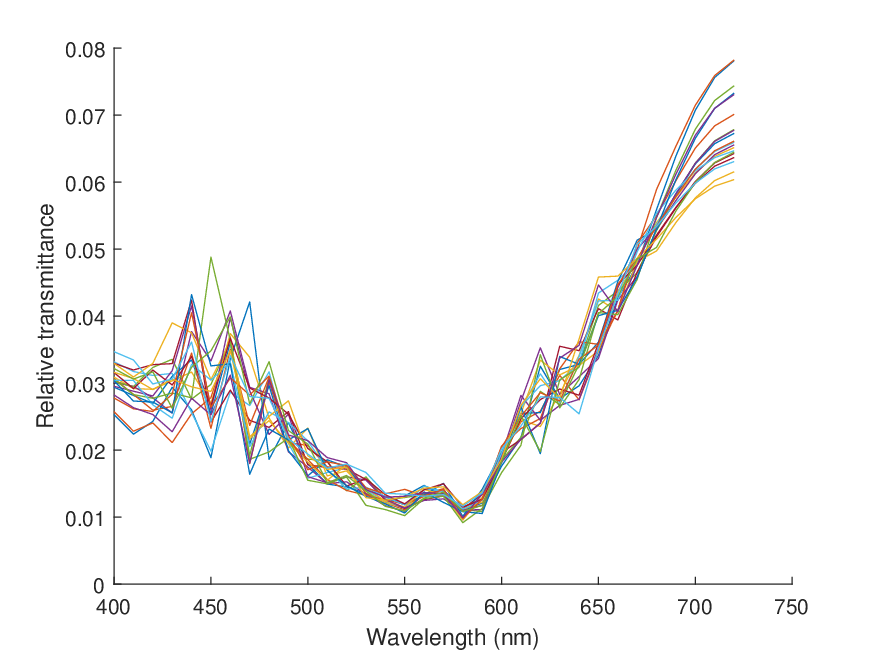}
\caption{The first 20 of N filter estimates (from Observer-I's data)}
  \label{Figure:FilterEstimates_20}
\end{figure}

Fig.~\ref{Figure:FilterEstimates_20} displays the first 20 of N (where N = 1000) filter estimates (for Observer-I) derived from the bootstrapped filter estimation process. This visualization highlights the variability in the estimated filter characteristics, showcasing the range of possible filter transmittance vectors, $\hat{\textbf{f}}^{(i)}$, that could transform the CMFs without the filter to those with the filter, within the bounds of experimental noise. 

\begin{figure}[h]
\centering\includegraphics[width=10cm]{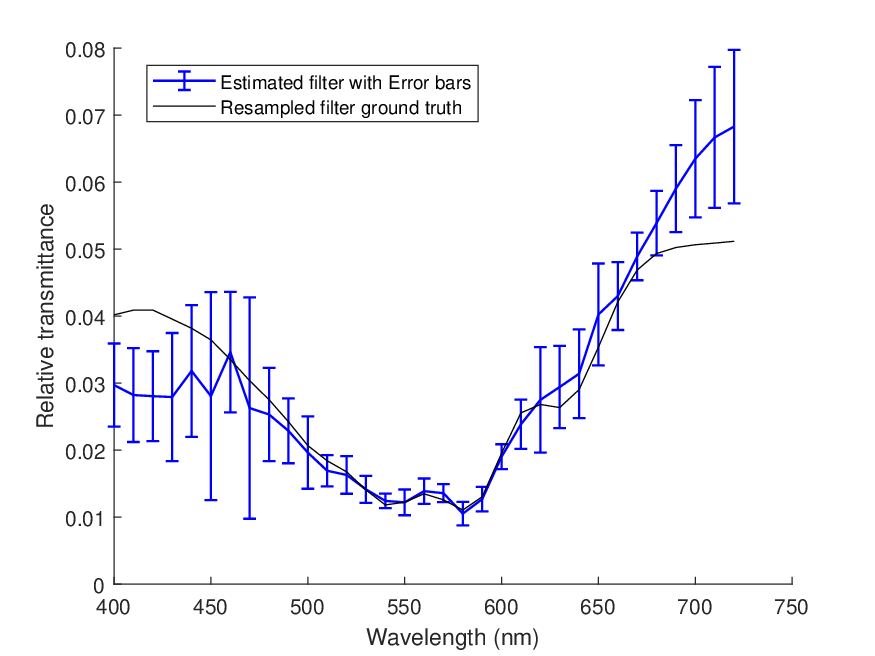}
\caption{Observer-I : Mean estimated filter and the resampled ground truth of the filter, error bars representing 2 $\sigma_\mathbf{f}(\lambda)$}
  \label{Figure:EstimatedFilter_Groundtruth}
\end{figure}

\begin{figure}[h]
\centering\includegraphics[width=10cm]{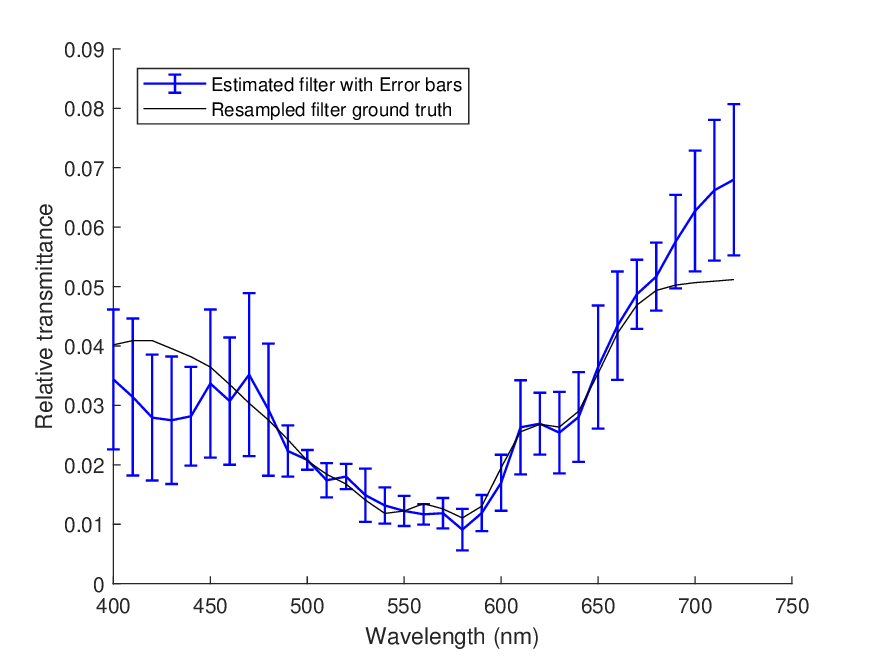}
\caption{Observer-C : Mean estimated filter and the resampled ground truth of the filter, error bars representing 2 $\sigma_\mathbf{f}(\lambda)$}
  \label{Figure:EstimatedFilter_Groundtruth_Cristina}
\end{figure}

The mean estimated filter, along with error bars representing two times the standard deviation ($2\sigma_\mathbf{f}$) of the filter estimate at each wavelength from 400~nm to 720~nm, at 10~nm intervals, is presented in Fig.~\ref{Figure:EstimatedFilter_Groundtruth} for Observer-I and in Fig.~\ref{Figure:EstimatedFilter_Groundtruth_Cristina} for Observer-C. This figure allows for a direct comparison with the ground truth filter SPD measurement, providing insight into the accuracy and reliability of the filter estimation algorithm. It is worth noting that each estimated filter transmittance is normalized to have a sum of 1, as only the shape of the filter's transmittance curve can be estimated, not the absolute transmittance. Similarly, the resampled ground truth of the "purple" filter transmittance is also normalized in this way to facilitate comparison.

\section{Discussion}

\subsection{Performance of Filter estimation algorithm}

The experimental results validate the hypothesis that a spectral filter (and a linear transformation matrix) can be computed to convert unfiltered CMFs to those obtained with a filter. The estimated filter's spectral transmittance closely aligns, at least in the central region, with the measured transmission properties of the physical filter used in the experiment, for each observer. This alignment underscores the robustness of the filter estimation algorithm in capturing the spectral characteristics of physical filters used in color-matching experiments. It can be observed that under 440~nm and above 690~nm, the estimated filters is no longer in agreement with the ground truth. This is potentially due to reduced illumination induced by filtering during the color-matching experiments. The observers did report that matching is harder at the edges of the visible spectrum under the filtered experimental condition, which is also evident given the larger error bars of the estimated filter in those regions. 

\subsection{Transformation between two different CMFs sets}

\begin{figure}[ht]
\centering\includegraphics[width=13cm]{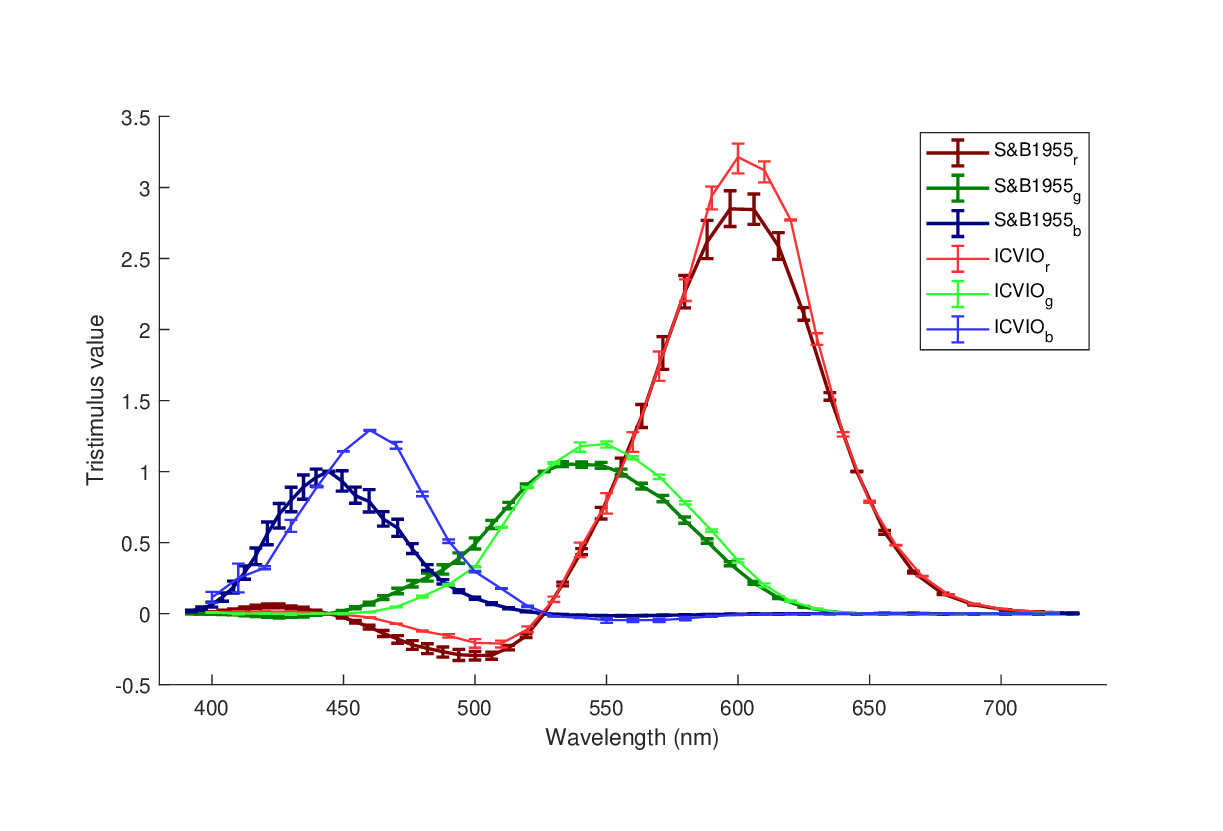}
\caption{ICVIO mean CMFs vs Stiles and Burch 1955 mean CMFs}
  \label{Figure:meanICVIO_vs_meanSB1955}
\end{figure}

Given the promising results of the filter estimation algorithm, we decided to compute the spectral filter and $3 \times 3$ matrix required to transform between the mean CMFs from two different datasets. The CMFs chosen for this exercise are the Stiles and Burch 1955 2$^\circ$ mean  observer CMFs, and the 2$^\circ$ mean observer CMFs from our previously published work\cite{ragoo2024apparatus}, which we are calling the ICVIO mean observer CMFs here. The two CMFs sets are plotted in Fig. \ref{Figure:meanICVIO_vs_meanSB1955}. 

\begin{figure}[h]
\centering\includegraphics[width=13cm]{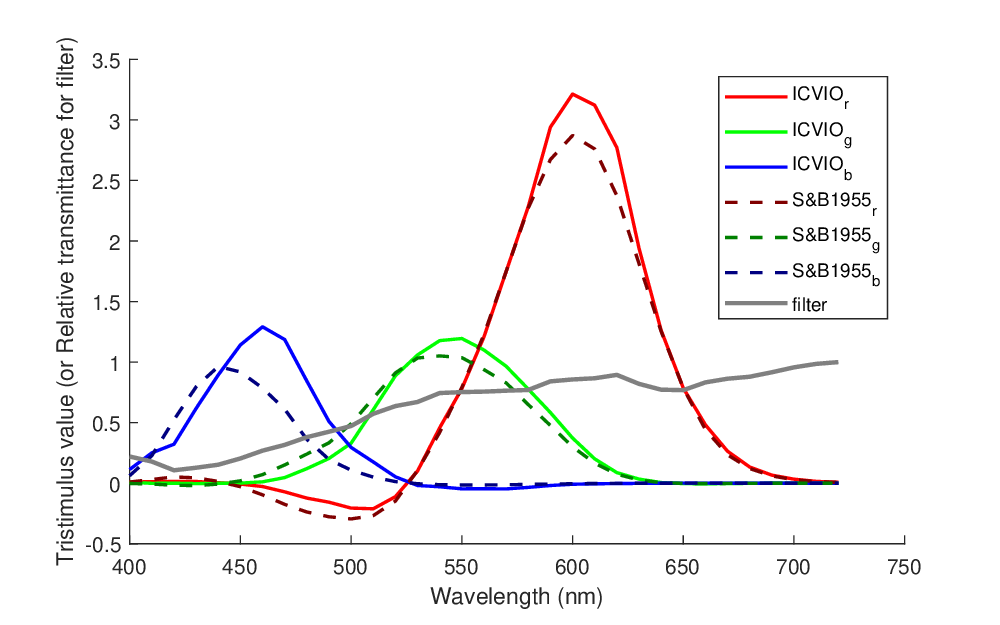}
\caption{The filter that would transform from the Stiles and Burch 1955 CMFs (dotted) to that of ICVIO's}
  \label{Figure:meanICVIO_vs_meanSB1955_wFilter}
\end{figure}

The filter that would convert from the Stiles and Burch 1955 2$^\circ$ mean observer CMFs to that of the ICVIO mean observer CMFs is shown in Fig. \ref{Figure:meanICVIO_vs_meanSB1955_wFilter}. The computed filter exhibits a distinctive spectral profile, with progressively decreasing transmittance toward shorter wavelengths and relatively higher transmittance in the middle and long wavelength regions. This spectral shape closely resembles that of a "yellow" filter, with significant attenuation in the blue region of the spectrum. This finding is particularly noteworthy as it aligns with well-documented physiological changes that occur in the human visual system with age—specifically, the yellowing of the crystalline lens\cite{pokorny1987aging}. The lens naturally accumulates yellow pigmentation over time, which selectively absorbs short-wavelength light and alters color perception, particularly in the blue region of the spectrum. This result was expected, given that the mean age of the observers comprising the ICVIO dataset is 49 years old, compared to mean age of 30 years old for observers in the Stiles and Burch 1955 dataset. This adds weight to our previous hypothesis~\cite{ragoo2024apparatus} that the discrepancy observed between the Stiles and Burch 1955 2$^\circ$ mean observer CMFs and our own ICVIO mean observer CMFs, can be attributed, at least partially, to physiological factors such as lens aging.

As a sanity check to verify that the filter and the $3 \times 3$ matrix computed would indeed result in the ICVIO 2$^\circ$ mean observer CMFs when applied to the Stiles and Burch 1955 2$^\circ$ CMFs, an estimate of the ICVIO CMFs is thus computed :

\begin{equation}
    \overline{\mathbf{C}}_\mathrm{ICVIO} = \mathrm{diag}(\mathbf{f}_{\mathrm{ICVIO}})\mathbf{C}_\mathrm{SB1955}\mathbf{M}_\mathrm{ICVIO}^{-1} 
\end{equation}

where :
\begin{itemize}
    \item $\overline{\mathbf{C}}_\mathrm{ICVIO}$ are the reconstructed ICVIO 2$^\circ$ mean observer CMFs. 
    \item $\mathbf{f}_\mathrm{ICVIO}$ is the computed spectral filter that would convert from the Stiles and Burch 1955 2$^\circ$ mean observer CMFs to that of ICVIO's
    \item $\mathbf{C}_\mathrm{SB1955}$ are the Stiles and Burch 1955 2$^\circ$ mean observer CMFs
    \item $\mathbf{M}_\mathrm{ICVIO}$ is the computed $3 \times 3$ linear transformation matrix
\end{itemize}

\begin{figure}[ht]
\centering\includegraphics[width=13cm]{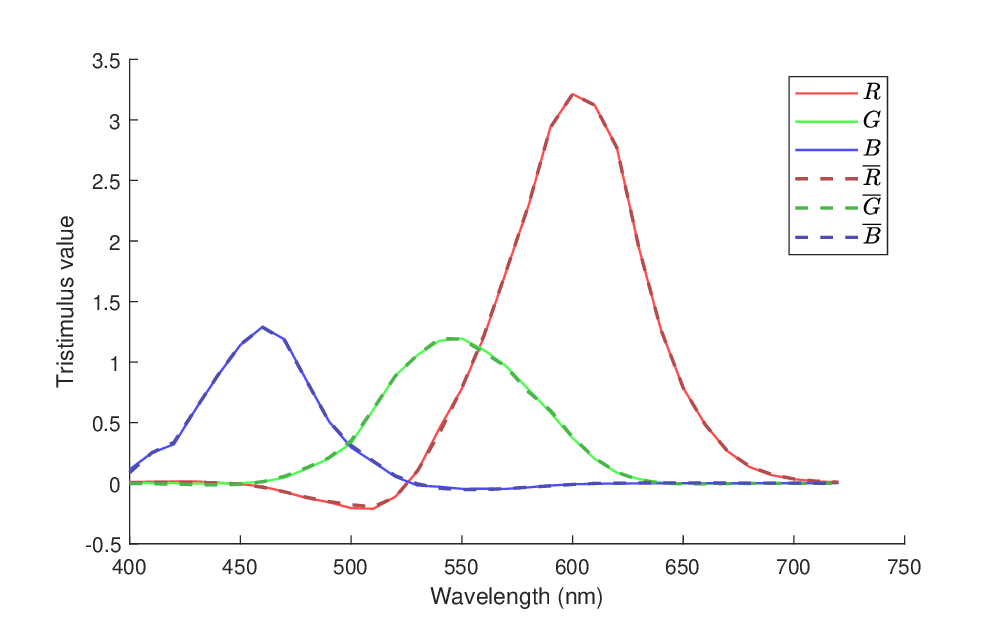}
\caption{Mean Observer CMFs vs reconstructed mean observer CMFs when estimated filter and 3x3 matrix is applied to Stiles and Burch 1955 CMFs}
  \label{Figure:meanICVIO_vs_reconstructed_ICVIO}
\end{figure}

The reconstructed ICVIO mean observer CMFs are compared to the originals in Fig. \ref{Figure:meanICVIO_vs_reconstructed_ICVIO}. There is an almost perfect agreement (well within the matching uncertainty) between the reconstructed ICVIO 2$^\circ$ mean observer CMFs, $\overline{\mathbf{C}}_\mathrm{ICVIO}$, and the original ICVIO 2$^\circ$ mean observer CMFs, $\mathbf{C}_\mathrm{ICVIO}$, which suggests that almost all the difference between the Stiles and Burch 1955 2$^\circ$ CMFs and original ICVIO 2$^\circ$ mean observer CMFs is modeled effectively by the estimated spectral filter and the linear transformation matrix. 

\subsection{Implications for Observer Variability}

The results of this study demonstrate the practical utility of the proposed method in modeling inter-observer differences in color-matching functions. By representing individual CMFs as a combination of a spectral filter and a $3 \times 3$ linear transformation matrix, the dimensionality of individual CMF representation is significantly reduced. This approach eliminates the need to record individual CMFs as three separate $N \times 1$ vectors (for red, green, and blue CMFs). Individual differences can instead be effectively represented by a single spectral filter that transforms an individual observer’s CMFs to correspond to a standard CMF set.

This reduction in dimensionality has several important implications:

\begin{itemize}
    \item \textbf{Efficient Representation of Individual Differences:}  
    The ability to model individual CMFs using a single spectral filter simplifies the characterization of inter-observer variability. The spectral filter captures physiological differences, such as lens pigmentation or macular pigment density, while the linear transformation matrix accounts for systematic variations in color-matching behavior. This framework provides a unified and efficient way to quantify observer-specific deviations from standard CMFs.

    \item \textbf{Potential for Shortened Color-Matching Experiments:}  
    Since fewer parameters are required to model individual CMFs, it is plausible that fewer color-matching data points may suffice to estimate an individual's spectral filter. This could lead to shorter experimental protocols while maintaining high accuracy in reconstructing individual CMFs. Such efficiency gains are particularly valuable in large-scale studies or applications requiring rapid assessments of observer variability.

    \item \textbf{Insights into Physiological Sources of Variability:}  
    By linking spectral filters to known physiological factors—such as age-related yellowing of the lens—the method provides a practical tool for studying how these factors influence inter-observer variability. For instance, the yellowing effect observed in this study aligns with previous findings on lens aging and its impact on short-wavelength light absorption.
\end{itemize}

\subsection{Limitations}

\subsubsection{Experiment and optimization method}
The experimental scope of this study presents a notable limitation, as it involved only two observers performing a color-matching experiment to derive color-matching functions (CMFs) with and without a physical filter. Given the demographic differences between the two observers, this constrained setup was sufficient to validate the method's ability to estimate a spectral filter that transforms between the two sets of CMFs, successfully recovering the transmittance properties of the physical filter to some extent. However, it remains uncertain whether the approach would perform similarly for other observers. The limited amount of observers limits the generalizability of the findings and raises questions about the robustness of the methodology across individuals. Future experiments involving a broader range of observers would be essential to evaluate whether the method consistently captures the physical properties of filters across different observers.

Another limitation lies in the constraints of the filter estimation algorithm itself. The algorithm estimates only the relative shape of the spectral transmittance curve rather than its absolute magnitude, necessitating normalization for comparisons with ground truth data. This reliance on normalization introduces additional dependencies that could influence the accuracy of the estimated filters. While bootstrapping methods were employed to quantify uncertainty in filter estimation, they primarily reflect variability in repeated measurements and do not address deeper issues such as model bias or parameter sensitivity. Furthermore, optimization techniques could be refined to improve stability and accuracy under varying experimental conditions. For example, k-fold cross-validation could be used to determine optimal regularization parameters, such as the $\alpha$ smoothing parameter, ensuring that the model achieves an appropriate balance between flexibility and robustness.

As we saw in the experimental section, our optimization method works rather well for our purposes. However, there are other algorithms in this space which are relevant here. In \cite{FinlaysonZhu} Finlayson and Zhu present an alternating least-squares (ALS)  algorithm which also incorporates a smoothness constraint. The ALS algorithm works iteratively. First the best filter is found (for a given initial linear transform) then this is held fixed and the best linear transform is solved for. The linear transform is then held fixed and a new filter is found. The algorithm iterates like this until convergence. The ALS approach has the advantage that for a given initialisation condition it will always minimize the fitting error though the process may not necessarily converge to a global optimum. However, also in \cite{FinlaysonZhu} it is shown how multiple initialisation conditions might be used to find the best overall result. Relatedly, in \cite{FinlaysonFarup} the optimal transform and filter is found by warping the chromaticity spaces of the two sensor sets. Here a global optimum is found.

In the literature, the method of Riverz \cite{Rivertz} is the most similar to Equations (\ref{Eq_optimisationProblem}) through (\ref{Eq_regQuadraticProblem}). There, his {\it simplified} method has a similar formulation to the one we derived above, it too is a null-space method. Rivertz applies his null-space method to the problem of making camera sensors colorimetric (a filter and linear transform is found that together make a particular camera better able to make colorimetric measurements). However, that study found that the ALS method worked better for a mean-square error metric. Rivertz's null-space method was found be a more competitive optimisation  when the Vora-Value\cite{Vora} metric is used to assess similarity to the human visual system's colour matching functions. With respect to the Vora-Value, the ALS and Rivertz's simplified method delivered similar levels of performance. However, the ALS method was recently modified so it could optimise for the Vora Value \cite{FinlaysonZhuVora}. Here, ALS always works better than the  null-space method.

In spite the caveats above, we used our null-space formulation because it satisfactorily accounts for our data.

\subsubsection{Modeling inter-observer variability}

The implications of this work for modeling inter-observer variability are significant but require careful consideration of the underlying assumptions. The study demonstrates that the filter estimation algorithm effectively "summarizes" the difference between two unrelated CMF datasets by estimating a spectral filter and a linear transformation matrix. This is evident in Fig. 8, where the reconstructed ICVIO CMFs—obtained by applying the computed filter and transformation matrix to the SB1955 CMFs—show near-perfect agreement with the original ICVIO dataset. The estimated filter, resembling a "yellow" filter, aligns with the hypothesis that age-related lens yellowing contributes to differences between the datasets. However, while this result supports the feasibility of using such filters to model dataset differences, it may also obscure other underlying causes, such as physiological factors beyond optical filtering.

The current approach assumes that differences between CMF datasets can be attributed primarily to optical filtering effects, such as lens yellowing, while other physiological factors remain unmodeled. For instance, variations in macular pigment density, cone photopigment optical densities, and spectral shifts in cone photopigment sensitivity—all parameters described in Stockman’s cone fundamentals\cite{cie2006fundamental,stockman2000spectral}—are known to influence color perception but are not explicitly incorporated into this framework. These factors could interact with optical filtering effects in complex ways that are not captured by a single spectral filter and linear transformation matrix. Future work should aim to integrate these physiological parameters into the model to provide a more comprehensive understanding of inter-observer variability in CMFs and disentangle the contributions of different factors to observed dataset differences.



\section{Conclusion}

This study set out to determine whether color-matching data could be used to recover the transmittance properties of a physical filter placed in the optical path. The results demonstrate that the filter estimation algorithm successfully recovered these properties, showing good agreement with measured data in central wavelength regions. However, deviations at edges of the visible spectrum were observed, likely due to reduced illumination and experimental noise. It is important to note that these results reflect relative transmittance rather than absolute values. Further testing with a broader observer pool is necessary to evaluate whether the method remains robust in estimating filter transmittance across varying individuals.

Beyond physical filter characterization, this research explored modeling inter-observer variability by capturing differences between CMF datasets in a simplified manner. The transformation between the Stiles and Burch 1955 and ICVIO CMF datasets demonstrated that these differences can be elegantly represented using a computed spectral filter and a linear transformation matrix, effectively summarizing individual CMFs without requiring $N \times 3$ functions. This dimensionality reduction offers significant practical advantages in characterizing observer-specific color-matching functions.

Future work should focus on incorporating a wider range of physiological parameters in the existing model for a more complete picture of inter-observer variability. Additionally, increasing the observer pool and refining experimental and optimization methods will enhance generalizability to the broader population. These efforts will provide deeper insights into individual variability in color perception and further strengthen the framework developed in this study.

Lastly, building on the apparatus detailed in our previous publication\cite{ragoo2024apparatus}, this study demonstrates that despite experimental noise and changes in conditions—such as the addition of a filter in front of the bipartite field—the observer was able to perform reliable color-matching. The results effectively captured the impact of the spectral filter on color-matching behavior, highlighting the stability of the apparatus in accommodating such modifications. While noise was present, the experimental setup proved robust enough to facilitate meaningful measurements, supporting its continued use in color vision studies.  


\begin{backmatter}

\bmsection{Funding}
This research was funded by the Research Council of Norway over the project ‘Individualised Colour Vision-based Image Optimisation’, grant number 287209.

\bmsection{Acknowledgments}
This research was funded by the Research Council of Norway over the project ‘Individualised Colour Vision-based Image Optimisation’, grant number 287209.

\bmsection{Disclosures}
The authors declare no conflicts of interest.

\bmsection{Data availability}
Data underlying the results presented in this paper are not publicly available at this time but may be obtained from the authors upon reasonable request.

\end{backmatter}


\bibliography{main}

\end{document}